\documentstyle[twocolumn,aps,floats]{revtex}

\begin{document}

\draft \tolerance = 10000

\setcounter{topnumber}{1}
\renewcommand{\topfraction}{0.9}
\renewcommand{\textfraction}{0.1}
\renewcommand{\floatpagefraction}{0.9}
\newcommand{\br}{{\bf r}}

\twocolumn[\hsize\textwidth\columnwidth\hsize\csname
@twocolumnfalse\endcsname

\title{ Newton Equations May Be Treated as Diffusion Equations in the Real Time and
Space Fields of Multifractal Universe (Masses are Diffusion Coefficients
of Diffusion-like Equations}
\author{L.Ya.Kobelev\\
 Department of  Physics, Urals State University \\ Lenina Ave., 51,
Ekaterinburg, 620083, Russia  \\ E-mail: leonid.kobelev@usu.ru} \maketitle

\begin{abstract}
In thirties years of last century Dirac proposed to treat Shr\"{o}dinger
equation as the equation of diffusion with imaginary diffusion
coefficient. In the frame of multifractal theory of time and space ( in
this model our the multifractal universe is consisting of real time and
space fields) in the works \cite{kob1}-\cite{kob16} was analyzed how the
fractional dimensions of real fields of time and space influence on
behavior of different physical phenomena. In this paper the Newton
equations of the multifractal universe (considered for the first time in
\cite{kob1}- \cite{kob3}) are generalized and is treated as the equations
of diffusion with mass of bodies (depending of fractional dimension of
place, where these bodies located) as a coefficient of diffusion. The
realization of this point of view for inhomogeneous time equations ( the
analogies of Newton equations) is carried out too. The last lead to
introducing new sort of masses: the masses that characterize the inertia
of inhomogeneous time flows with space coordinates changing.
\begin{center}
CONTENTS:
\end{center}
1. Introduction\\2. Newton Equations in the Multifractal Universe
\\3. Generalized Newton Equations and Its Diffusion Interpretation
\\4. Generalized Inhomogeneous Time Equation and Its Diffusion
Interpretation\\5. Conclusions
\end{abstract}

 \vspace{1cm}

]

\section {Introduction}

This paper based on some concepts of physical models of the theory of
multifractal time and space which are consequences of this theory (Kobelev
\cite{kob1}- \cite{kob16}). The multifractal model of space and time
treats the time and the space with fractional dimensions as real fields.
Universe is formed only by these fields , i.e.our universe is the
fractional real material time and fractional real material space. All
other fields are born by the fields of time and space by means of their
fractional dimensions. As the time and the space are material fields with
fractional dimensions and multifractal structure (multifractal sets,) they
are defined on the sets of their carriers of measure (physical vacuum for
our universe which born her when "big-bang" happened). In each the time
(or the space) point ( "points" are approach for very small "intervals" of
time or space and "intervals" are the multifractal sets with global
dimensions for their sets, that play role of local dimensions for universe
in whole) the dimensions of time (or space) determine the densities of
Lagrangians energy for all physical fields ( or new physical fields for
space) in these points. Time and space are binding by relation $d{\bf
r}^{2}-c^{2}dt^{2}=0$ (this relation is only a good approach, more precise
relations see at \cite{kob2}). The purpose of this paper is consideration
of very interesting problem ( in the frame of mathematical formalism of
the multifractal model of time and space introduced in the papers of
Kobelev and presented in \cite{kob1}- \cite{kob16}): may masses of bodies
be treated as characteristic of the real space that are analogies of the
diffusion  coefficient ? Such interpretation will be the analogy of Dirac
interpretation of Shr\"{o}dinger equation. In that case main equations of
modern physics (Newton and Shr\"{o}dinger equations) are the analogy of
diffusion equation and their nature caused by diffusion processes in the
real time and space fields of our multifractal universe.

\section {Newton equations in the multifractal universe}

Newton equations in the multifractal universe considered in
\cite{kob1}-\cite{kob3}. These equations read
\begin{equation}    \label{1}
D_{-,t}^{d_{t}(r,t)}[mD_{+,t}^{d_{t}(r,t)} {\bf r}(t)]=
D_{+,r}^{d_{r}}[m\Phi_{g}({\bf r}(t))]
\end{equation}
\begin{equation}    \label{2}
D_{-,r}^{d_{r}}D_{+,r}^{d_{r}}\Phi_{g}({\bf r}(t))+
\frac{b_{g}^{2}}{2}\Phi_{g}({\bf r}(t))=\gamma
\end{equation}
In the Eq.(\ref{2}) the constant $b_{g}^{-1}$ has order of  size of
universe and introduced with purpose to extend the class of functions on
which the generalized fractional derivatives concept is applicable. These
equations do not present a closed system because of the fractionality of
spatial dimensions. Therefore we approximate the fractional derivatives
with respect to space coordinate as $D_{+,{\bf r}}^{d_{{\bf r}}}\approx
{\bf \nabla}$, i.e. approximate by usual space derivatives. In the Eqs.
(\ref{1})-(\ref{2}) is used the integral functionals $D_{+,t}^{d_{t}}$
(both left-sided and right-sided) which are suitable to describe the
dynamics of functions defined on multifractal sets (see
\cite{kob1}-\cite{kob3}). These functionals are simple and natural
generalization of the Riemann-Liouville fractional derivatives and
integrals and read:
\begin{equation} \label{3}
D_{+,t}^{d}f(t)=\left( \frac{d}{dt}\right)^{n}\int_{a}^{t}
\frac{f(t^{\prime})dt^{\prime}}{\Gamma
(n-d(t^{\prime}))(t-t^{\prime})^{d(t^{\prime})-n+1}}
\end{equation}
\begin{equation} \label{4}
D_{-,t}^{d}f(t)=(-1)^{n}\left( \frac{d}{dt}\right)
^{n}\int_{t}^{b}\frac{f(t^{\prime})dt^{\prime}}{\Gamma
(n-d(t^{\prime}))(t^{\prime}-t)^{d(t^{\prime})-n+1}}
\end{equation}
where $\Gamma(x)$ is Euler's gamma function, and $a$ and $b$ are some
constants which take values from interval $(0,\infty)$. In these
definitions, as usually, $n=\{d\}+1$ , where $\{d\}$ is the integer part
of $d$ if $d\geq 0$ (i.e. $n-1\le d<n$) and $n=0$ for $d<0$. If $d=const$,
the generalized fractional derivatives (GFD) (\ref{3})-(\ref{4}) coincide
with the Riemann - Liouville fractional derivatives ($d\geq 0$) or
fractional integrals ($d<0$). When $d=n+\varepsilon (t),\, \varepsilon
(t)\rightarrow 0$, GFD can be represented by means of integer derivatives
and integrals. For $n=1$, i.e. $d=1+\varepsilon$, $\left| \varepsilon
\right| <<1$ it is possible to obtain:
\begin{eqnarray}\label{5}
D_{+,t}^{1+\varepsilon }f({\mathbf r}(t),t)\approx
\frac{\partial}{\partial t} f({\mathbf r}(t),t)+ \nonumber \\ +
a\frac{\partial}{\partial t}\left[\varepsilon (r(t),t)f({\mathbf
r}(t),t)\right]+ \frac{\varepsilon ({\mathbf r}(t),t) f({\mathbf
r}(t),t)}{t}
\end{eqnarray}
where $a$ is a $constant$ and determined by  choice of the rules of
regularization of integrals (\cite{kob1}-\cite{kob2}, \cite{kob7}) (for
more detailed see \cite{kob7}) and the last member in the right hand side
of (\ref{5}) is very small. The selection of the rules of regularization
that gives a real additives for usual derivative in (\ref{3}) yields
$a=0.5$ or $a=1$ for $d<1$ \cite{kob1}. The functions under the integral
sign in the (\ref{3})-(\ref{4}) we consider as the generalized functions
defined on the set of the finite or Gelfand functions \cite{gel}. The
notions of the GFD, similar to (\ref{3})-(\ref{4}), can also be defined
and for the space variables ${\mathbf r}$. The definitions of GFD
(\ref{3})-(\ref{4}) need in the connections between the fractal dimensions
of time $d_{t}({\mathbf r}(t),t)$ and the characteristics of physical
fields (say, potentials $\Phi _{i}({\mathbf r}(t),t),\,i=1,2,..)$ or
densities of Lagrangians $L_{i}$) and such connections were defined in the
cited works. Following \cite{kob1}-\cite{kob15}, we define this
connections by the relation
\begin{equation} \label{6}
d_{t}({\mathbf r}(t),t)=1+\sum_{i}\beta_{i}L_{i}(\Phi_{i} ({\mathbf
r}(t),t))
\end{equation}
where $L_{i}$ are densities of energy (Lagrangian densities) of physical
fields, $\beta_{i}$ are dimensional constants with physical dimension of
$[L_{i}]^{-1}$ (it is worth to choose $\beta _{i}^{\prime}$ in the form
$\beta _{i}^{\prime }=a^{-1}\beta _{i}$ for the sake of independence from
the regularization constant and select the $\beta=Mc^{2}$ where $M$ is the
mass of the body that born considered gravitational field). The definition
of the time as the system of subsets and definition of the FD for $d_{t}$
(see (\ref{6})) connects the value of fractional (fractal) dimension
$d_{t}(r(t),t)$  with each time instant $t$. The latter depends both on
time $t$ and coordinates ${\mathbf r}$. If $d_{t}=1$ (an absence of
physical fields) the set of time has topological dimension equal to unity.
For large energy densities (e.g.,for cases when gravitational field is
large (the domain of space where $r<r_{0}$) Eqs.(\ref{1})-(\ref{2})
contain no divergencies \cite{kob1} since integral-differential operators
of the generalized fractional differentiation are reduced to the
generalized fractional integrals (see (\ref{1})). We bound the
consideration only by the case when relations
$d_{t}=1-\varepsilon({\mathbf r}(t),t))$, $|\varepsilon|\ll 1$ are
fulfilled.  In that case the GFD (as was shown in \cite{kob7} ) may be
represented (as a good approach) by ordinary derivatives and relations
(\ref{1}), (\ref{5}) are valid.  Now we can determine the $d_{t}$ for
distances much larger than the gravitational radius $r_{0}$ (for the
problem of a body motion in the field of spherical-symmetric gravitating
center) as (see \cite{kob1} for more details)
\begin{equation} \label{7}
d_{t}\approx 1+\beta_{g}\Phi_{g}+\beta \Phi_{m}
\end{equation}
where $\Phi_{g}$ is the gravitational potential of mass $M$ and
$\Phi_{m}$ is the gravitational potential born by the body with mass $m$
in its center. So the equations (\ref{1})-(\ref{2}) reed ( we used for GFD
approach of (\ref{5})\cite{kob16})
\begin{equation}\label{8}
  [1-2\varepsilon({\bf r(t)}]\frac{d^{2}}{dt^{2}}{\bf r}= {\bf F}_{g}
\end{equation}
where
\begin{equation}\label{9}
  {\bf F}_{g}= -{\bf \nabla}\frac{\gamma M}{r} ,
  \varepsilon=\beta_{g}\Phi_{g}+\beta_{m} \Phi_{m}
\end{equation}
We had neglected by  the fractional parts of spatial dimensions and by the
contributions from the term with $b_{g}^{-1}$. Now we take the $\beta_{m}
=\beta_{g}=c^{-2}$ for potentials (or $\beta=(Mc)^{-2}$ for Lagrangian
density $L$). Than equation (\ref{8}) gives the small corrections to
Mercury perihelium rotations of general relativity (see \cite{kob17}). It
is the example of limited character of general relativity principle of
equivalence.

 \section{Generalized Newton Equations and their diffusion interpretation}

In this paragraph we write down  the generalization of modified Newton
equations (\ref{1})-(\ref{2}) in the multifractal time universe. Let the
gravitational forces only is presence . This generalization based on a
quite natural assumption: if the every point of the real spatial field
determines their fractional ( temporal and spatial) dimensions, the masses
in this point (as the one of characteristics of the real spatial field)
must have dependence of the fractional dimensions in this point (i.e. at
$d_{t}$ and $d_{r}$ ). So, the Eqs. (\ref{1})-(\ref{2}) will have the form
\begin{equation}\label{10}
D_{-,t}^{d_{t}(r,t)}[m_{d_{t},d_{r}}D_{+,t}^{d_{t}(r,t)} {\bf r}(t)]=
D_{+,r}^{d_{r}}[m_{d_{t},d_{r}}\frac{\gamma M}{r}]
\end{equation}
If the mass of accelerated body $m_{d}$ has no dependencies of fractional
dimensions the Eq.(\ref{10}) and Eqs. (\ref{1})-(\ref{2}) are coinciding.
We consider now the simple method of receiving of the equation (\ref{10})
from qualitative reasoning. Let $E_{r}$ is the density of energy at the
point ${\bf r}$ of real space field. On the one hand the gradient of it
gives for current ${\bf j}$ equation
\begin{equation}\label{11}
 {\bf j_{E}} \sim {\nabla}E(r)
\end{equation}
On another hand this current is proportional to changing ( with changing
of time) of gradient $E(r)$ in the temporal space
\begin{equation}\label{12}
{\bf j_{E}} \sim \frac{\partial^{2}}{\partial t^{2}}E(r)
\end{equation}
The latter equation describes the diffusion of the energy at space point
$r$ in the temporal space under the influence of gradient of it. So
designate a diffusion coefficient in temporal space as $m_{d_{t},d_{r}}$.
We may compare now the (\ref{12}) to the (\ref{11}) and separate $E$ in
the (\ref{13}) into the coefficient of diffusion $m$ and the space field
transferring the energy ${\bf r}$ and replace $E$ by gravitational energy
in the (\ref{11}). We receive the Newton equations in the form (\ref{10}).
Thus we have the interpretation of Newton equations as the equations are
describing the masses of bodies as the diffusion coefficients of
transferring of energy density of the real space field of the point ${\bf
r}$ at the temporal space (at the spatial point ${\bf r}$) . We consider
the case when it is possible to neglect by corrections of order $
r_{0}r^{-1}$ where $ r_{0}=\frac{2\gamma M}{c^{2}}$ is the gravitational
radius, in the generalized fractional derivatives and replaced them by
usual derivatives, then the Eq.(\ref{11}) reads
\begin{equation}\label{13}
   m_{d_{t}}\frac{\partial^{2}}{\partial t^{2}}{\bf r}+
    \frac{\partial m_{d}}{\partial t} \frac{\partial }{\partial t} {\bf r}=
    {\bf \nabla} \frac{mM\gamma}{r}
\end{equation}
In the Eq.(\ref{13}) there is the  additional member with derivatives $m$
with respect to $t$ . This member describes the change of mass $m$ with
time. As in the multifractal theory of time and space (see
\cite{kob1}-\cite{kob16}) there are no the constant values, the appearing
of this member is very natural because all physical values in multifractal
universe are changing in time. For $d_{t}=1+\varepsilon, \varepsilon <<1$
this member is very small and not essential.

\section{Generalized inhomogeneous time equations and their diffusion
interpretation}

In the multifractal universe the  time and the space are  real  and
inhomogeneous fields. So for the time $t({\bf r})$ that depends of the
spatial points of space field ${\bf r}$ there is the equation that was
found for the first time in the paper \cite{kob1}. If we are to take into
account the fractionality of spatial dimensions ($d_{x}\ne 1,\;d_{y}\ne
1,\;d_{z}\ne 1$) (see \cite{kob1}- \cite{kob3}), we arrive to a new class
of equations that describe "temporal" physical fields (we shall call them
the "temporal" fields) generated by the space with fractional dimensions.
These equations are quite similar to the corresponding equations that
appear due to fractionality of time dimensions (the latter were given
earlier). In Eqs.(\ref{8})-(\ref{10}) of paper \cite{kob3}, \cite{kob7} we
must take $x=\br,; \alpha=\br$ and fractal dimensions
$d_{{\br}}(t(\br),\br)$ will be described by the Eq.(\ref{12}) of this
paper with $t$ being replaced by ${\bf r}$. Thus for time $t(\br(t))$ and
potentials $\Phi_{g}(t(\br),\br)$ (analogous of gravitational  field) the
equations analogous to Newton's will read (here the spatial coordinates
play the role of time)
\begin{equation}\label{14}
D_{-,\br}^{d_{\br}(\br,t)}m_{r}D_{+,\br}^{d_{\br}(\br,t)}t(\br)=
D_{+,t}^{d_{t}}[m_{r}\Phi_{g}(t(\br))]
\end{equation}
\begin{equation}\label{15}
D_{-,t}^{d_{t}}D_{+,t}^{d_{t}}\Phi_{g}(t(\br))+
\frac{b^{2}_{gt}}{2}\Phi_{g}(t(\br))=\gamma_{r}
\end{equation}
The equation (\ref{14}) describes the changes of time born by existence of
fractionality of spatial dimensions. The $m_{r}$ is the analogy of mass
the $m_{t}$ in the ordinary Newton equations and the last usually is
treated as measure of inertia of bodies. The fundamental constant
$\gamma_{r}$ is the analogies of gravitational constant $\gamma$ in field
of time. So the $ m_{r}$ may be treated as "measure of temporal inertia"
when the inhomogeneous time flows propagate through different spatial
domains.\\ Thus there are two sorts of "masses" in the multifractal
universe: the masses $m_{t}$ and the masses $m_{r}$. The first describe
the "bodies inertia" or diffusion of energy of the space with current of
time. The second masses $m_{r}$ describe the inhomogeneous of time flows
with changes of spatial places and may be treated as diffusion coefficient
of time energy propagation in the space of real field ${\bf r}$. So in the
multifractal universe the known conception of masses is doubled and it is
necessary to introduce the new sort of masses because there are exist not
only the space energy $2E_{t}=m_{t}[\frac{\partial {\bf r}}{\partial
t}]^{2}$ (non relativistic case), but the "time energy"
\begin{equation}\label{16}
 2E_{r}=m_{r}[\frac{\partial t}{\partial {\bf r}}]^{2}
\end{equation}
This relation characterize the time energy of real time field at point
$t({\bf r} )$. We may construct for this field the special relativity of
almost inertial systems in full analogy with papers \cite{kob4},
\cite{kob6}. For Lorentz transformation in the real time field than we
have
\begin{equation}\label{17}
 t'=\frac{t+v_{r}x}{\sqrt[4]{(1-\frac{v_{r}^{2}}{c_{r}^{2}})^{2}+4a_{0r}^{2}}},
\end{equation}
\begin{equation}\label{18}
x'=\frac{x+\frac{v_{r}t}{c_{r}^{2}}}{\sqrt[4]{(1-\frac{v_{r}^{2}}{c_{r}^{2}})^{2}+4a_{0t}^{2}}}
\end{equation}
In the Eqs. (\ref{17})-(\ref{18}) we use designation: $c_{r}=c^{-1}$,
$v_{r}=\frac{\partial t}{\partial x}$. The value $a_{0r}$ may be received
from correspondent the $a_{0}$ of paper \cite{kob4} by replacing the
fractional dimensions $d_{t}$ for $d_{r}$. For agreement with Lorentz
transformations respect to any moving in the real space it is necessary to
use relation: $v_{r}=\frac{\partial x}{\partial t}c^{-2}$. So there is
only two unknown values : the $m_{r}$, which is determine the time
diffusion (or "time inertia") of real time field in the multifractal
universe and temporal force $F(t)$ with unknown $\gamma_{r}$.

\section{conclusions}

The next main results is necessary to note:\\ 1.In this paper Dirac idea
about treating quantum mechanics as diffusion process (with imaginary
diffusion coefficient) is propagated for the domain of classical physics.
The last is possible only because of real nature of space and time fields
in the multifractal universe;\\ 2. the consideration of inhomogeneous time
equations (these equations are analogies of Newton equations for real time
field ) is based on consideration of the process of diffusion of "temporal
energy" in space when there are space coordinates changing. This energy
depends of fractional dimensions $d_{r}$ and when the last are changing
(that depends of changing of ${\bf r}$) with flow of time, the $d_{r}$ and
the temporal energy change too.\\3. Generalized Newton equations for the
multifractal universe is obtained;\\4.  Generalized temporal equation
(analogy of Newton equations for $t(\br)$) for multifractal universe is
obtained;\\5. The generalized temporal masses are presented (see also
\cite{kob1}-\cite{kob3})\\5. We pay attention once yet that if time and
space treat as convenient only marks for system of references (this point
of view contradicts the base assumption of the multifractal theory of
universe about the nature of time and space fields as the real fields)
diffusion interpretation of classic mechanics equations and nature of
usual and the temporal masses is impossible.

\end{document}